\documentclass[a4paper,11pt]{article}
\usepackage{pos}
\usepackage{natbib}
\title{Offline Signal Identification with GRANDProto300}
 \ShortTitle{Signal Identification with GP300}

\author*[a]{Pragati Mitra for the GRAND collaboration}

\affiliation[a]{Particles and Fundamental Interactions Division,Institute of Experimental Physics, University of Warsaw, Poland}

\emailAdd{pmitra@fuw.edu.pl}

\abstract{The GRANDProto300 (GP300) array is a pathfinder for the Giant Radio Array for Neutrino Detection (GRAND) project. Serving as a test bench, the GP300 array is expected to pioneer techniques of autonomous radio detection including identification and reconstruction of nearly horizontal cosmic-ray (CR) air showers, and shed light in understanding the interesting   `transition region'  from the galactic to extragalactic CR sources. An offline analysis of signal identification over ambient noise is crucial at this stage, where very relaxed self-triggering thresholds of radio antennas will be used for study purposes. In this work, we show  results and efficiency of  signal identification with classical approaches using a wide set of simulated realistic signal templates and also validated against measured background recorded by deployed prototypes. }

\ConferenceLogo{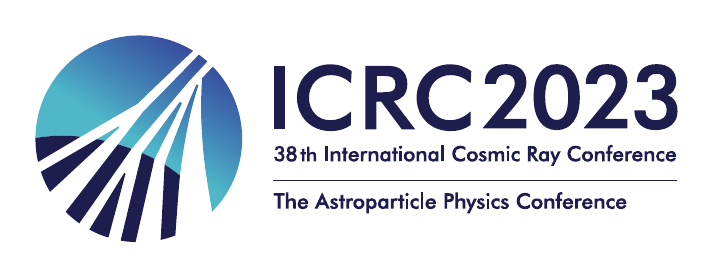}

\FullConference{%
38th International Cosmic Ray Conference (ICRC2023)\\
  26 July - 3 August, 2023\\
  Nagoya, Japan}

\begin{document}
\maketitle

\section{Introduction}
Detecting cosmic rays (CR) and neutrinos using radio emission at energies greater than 100 PeV  has advanced rapidly in the past decade. The Giant Radio Array for Neutrino Detection (GRAND)~\cite{torres, GRAND:2018iaj} will
employ 200,000 antennas, operating in self-triggering mode, to detect radio emission generated by extensive air showers (EASs),
that are initiated by ultra-high-energy (UHE) particles in the atmosphere. GRAND will consist of
roughly 20 separate, independent sub-arrays of approximately 10,000 radio antennas each, and
combined area of 200,000 km$^2$. It is being developed as a large-scale observatory with excellent sensitivity to observe UHE cosmic rays, neutrinos, and gamma rays. It is expected to provide novel insights into the origins of these particles.

The GRANDProto300 (GP300) experiment is the 300-antenna pathfinder stage of GRAND~\cite{Decoene:2019sgx}. 
This has the physics potential of measurement of cosmic-ray composition and energy measurement in the 10$^{16.5}$--10$^{18}$ eV range, focusing on the expected `transition region' in the CR energy spectra.  
The primary objectives of this stage include validating GRAND as a standalone radio detection array for EASs and optimizing self-triggering techniques. Operating as a self-triggering system is crucial for fully exploiting the potential of radio-detection technology with giant arrays. However, this poses a challenge due to the dominance of background sources such as high-voltage power lines, transformers, and airplanes, which generate transient radio signals in the tens-of-MHz frequency range. Nonetheless, promising results have been obtained regarding the feasibility of a self-triggering mode for EASs using radio detector has been tested by other experiments \cite{Ardouin:2010gz,Huege:2019ufo}. In the GP300 stage, algorithms will be developed to improve background rejection efficiency and identification of EAS events. More details on autonomous triggering for GRAND can be found in \cite{pablo, sandra}. Therefore, offline analysis  becomes crucial to developing and testing strong signal identification methods.

In this article, we present a classical approach to  signal identification based on template matching and discuss the effects of simulated and measured noise on the simulated CR pulses.

\section{Detector setup and Current status}
As of 2023, the deployment of prototypes for the GRAND experiment has commenced in multiple locations. Specifically, 13 units (known as GP13) have been deployed in Dunhuang, China as a part of GP300. Additionally, 4 units will be installed in Nanc\c{a}y, France, and 10 detection units are currently being deployed at the Pierre Auger Observatory site. Currently, the process of commissioning and collecting data is underway for these deployed detectors with the main objectives of understanding and improving upon the hardware responses, and characterizing the ambient background radio-frequency interference (RFI) at these sites.
The GRAND prototypes consist of butterfly antennas with three arms in the North-South (X), East-West (Y), and vertical (Z) directions,  operating in the 50-200~MHz frequency band. The height of the antennas is set to 3 m above ground to decrease the diffraction effect of radio waves off the ground.  The radio frequency (RF) receiving system of each detection unit of the GP13 comprises three parts: passive antenna probes, an RF chain, an analog-to-digital (AD) sampling system. More details about the hardware system can be found in \cite{xing}.

\section{Data and Simulations for Offline Analysis}
For this study, we have used a library of simulated CR radio signals produced with ZHAireS~\cite{zharies}.
This library contains a total of about 5000 showers containing proton and iron primaries, in energy range  0.02--4 EeV (uniform in logarithmic scale), and zenith  and azimuth ranging between 38$^{\circ}$--89$^{\circ}$ and 0$^{\circ}$--180$^{\circ}$, respectively. These simulations are performed at the geographical location of GP13. A star-shaped antenna layout is chosen, with the center on the shower axis. It has 8 arms and each arm has 20 antennas, giving a uniform coverage of the radio footprint on the ground. 

In order to produce realistic signals for signal identification, we fold antenna responses and simulated galactic noises to the simulated signal traces, along with a complete detector chain response that includes various stages of the Radio-Frequency (RF) receiving  system such as Linear Noise Amplifier (LNA), cables, Variable Gain Amplifier (VGA), filter, etc \cite{sandra}. 

\subsection{Template Matching}
\begin{figure}
\centering
\includegraphics[width=0.7\columnwidth]{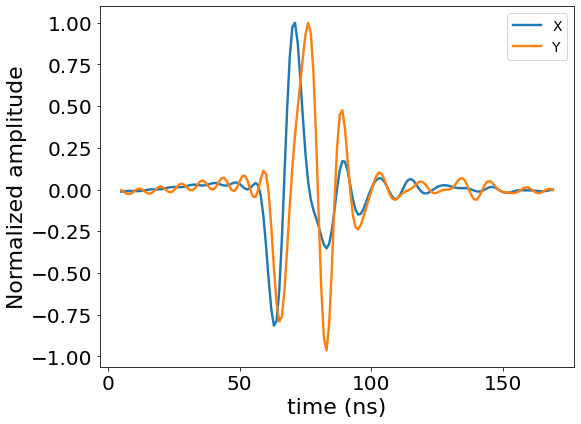}
\caption{CR templates obtained from averaging ZHAireS simulations and passing through antenna response for X, Y antenna channels.}
\label{template}
\end{figure}
Template matching is a widely used technique of extracting signal from noisy data based on the
convolution of a signal template $T$ with the time trace $u(t)$ ~\cite{Bezyazeekov:2019jbe}:
\begin{equation}
    S_{\mathrm{cc}}(t) = \sum_{t_{0}} T(t_0 -t) \, u(t) , \, A_{\mathrm{cc}} = \mathrm{max}(S_{\mathrm{cc}}) ,
\label{mf1}
\end{equation}
where A$_{\rm{cc}}$ is the maximum of the convolution, the square root of which defines the peak of the filtered signal $S_{\rm{cc}}$(t). Template matching is  effective for the detection of the signal position inside a trace significantly contaminated by stationary noise.

We construct the test signal template of length 160 ns  by averaging simulated CR pulses from ZHAireS, using the same library mentioned before, along with the antenna response (no noise or RF chain response added). For each simulation, we have considered 20 antennas with highest peak amplitudes. An example of the templates used for the X,Y channels of the antenna is shown in fig-\ref{template}.

\subsection {Application on Simulated Noise}
\begin{figure}

\includegraphics[width=0.5\columnwidth]{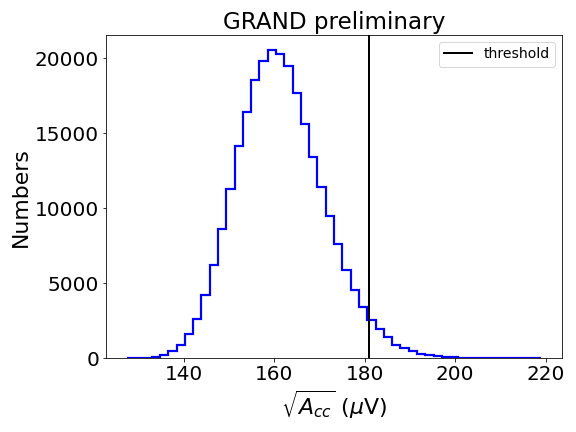}
\includegraphics[width=0.5\columnwidth]{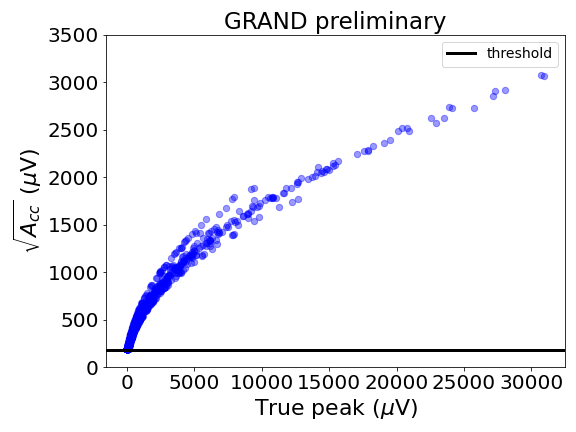}

\caption{\textbf{Left}: Distribution of $\sqrt{A_{\mathrm{cc}}}$ from template matching with simulated noise only traces. The threshold for CR detection is set from this plot at a 95\% CL, shown by the black line.
\textbf{Right}: Correlation between $\sqrt{A_{\mathrm{cc}}}$ and true peak amplitude of the simulated CR signals. The black line denotes the detection threshold shown in the figure on the left. Both the examples are shown for X channel, and the same procedure is done for channel Y.}
\label{scatter}
\end{figure}

Next, the template is matched with  the simulated CR signals with added simulated noise, and pulses that cross a detection threshold are counted.
The detection threshold is determined by matching the template with simulated noise-only traces containing simulated galactic noise and detector response. The distribution of 
$\sqrt{A_{\mathrm{cc}}}$ from the noise-only traces is shown in fig-\ref{scatter} (left). The threshold is set at a 2$\sigma$ level from the mean of this noise-only distribution, i.e., corresponding to a false-positive rate of 5\% at the antenna level. In reality, this number is to be tuned based on  hardware requirements, such as depending on the expected trigger rate. The correlation between the true signal peak and the filtered correlation for noisy CR signals (X channel)is shown in fig-\ref{scatter} (right). We find that the correlation is more prominent for stronger pulses. The filtered pulse peak is generally larger than the true peak due to the combined effect of added noise and RF chain.

We have studied the purity of this method, defined as the ratio of the number of noisy CR traces that pass the threshold to the total number of traces that pass the threshold. Here, the total number of traces consists of noisy CR traces with CR energies above 1 EeV and simulated noise traces. The purity is found to be about 85\% for both X and Y channels. 
We also have performed a detailed calculation of the efficiency of the entire simulated noisy CR sample considering coincidences, which is presented in Sec.3.3.

\subsection{Application on Measured Noise}
\begin{figure}
\centering
\includegraphics[width=0.7\columnwidth]{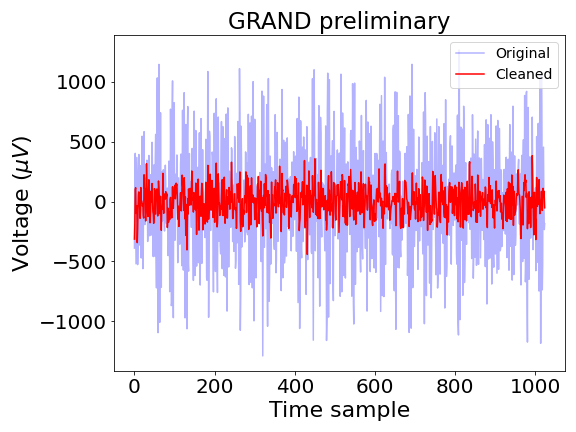}
\caption{ Noise measurement from GP13 data. Average noise traces before and after cleaning are shown for the X channel.}
\label{noise_profile}
\end{figure}

Since noise is an important factor for signal identification with radio arrays, it is crucial to test the performance of our signal characterization method with respect to the measured noise. For this purpose, we have utilized the data taken with  GP13 during the time period of 20-30 May 2023, which contains traces of a periodic mode with a 10-second interval. 
A detailed study of GP13 data can be found in \cite{pengxiong}. 
We have used an averaged profile for noise by collecting traces during night time observations for a particular detector unit. To ensure a stable noise contribution, traces with signal-to-noise ratio lower than 3 are selected, totaling about 3000 traces. In addition, the spectrum of these traces is processed to remove contamination of RFI spikes. First, outlying frequencies that are above a level of two times the rms of the spectrum are selected, and then  these are suppressed using a digital notch filter. The average time trace of measured noise, both before and after cleaning, is shown in fig-\ref{noise_profile}.
The average noise traces in X and Y channels are upsampled by a factor of 4 to match the sampling resolution of the simulated pulses and added to the simulated voltage traces. 
The template matching is performed on these noise-added traces, and the threshold for CR detection is determined from the correlation of the average noise trace with the template and the width of the distribution in fig-\ref{scatter} (left).

\begin{figure}
\centering
\includegraphics[width=0.7\columnwidth]{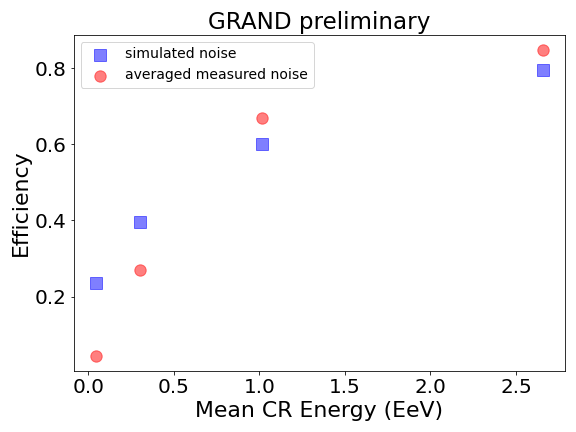}
\caption{Efficiency of signal identification as a function of CR energy for template matching with simulated (blue points) and measured noise (red points).}
\label{eff}
\end{figure}

Next, we study the efficiency of template matching for the two noise cases- simulated and measured. 
We accept a shower as a `CR pass' if at least 5 antennas are above the threshold in both X and Y channels. The efficiency is defined as the ratio of the number of true positives to the total number of CR showers,  and shown in fig-\ref{eff} for different bins of mean shower energy. It is seen that, for simulated noise the maximum efficiency reaches upto 60--80\% for the higher energy showers between 2--2.5 EeV. For the averaged measured noise, at the highest energies 
the efficiency is also ${\sim}80\%$. At the lower energy bins, the efficiency lies between 20--40 \% for the simulated noise case, while between 10--20 \% for  measured noise. 

It is important to note that the purpose of the study with the average measured noise is to validate the method for detecting CR pulses in the presence of generic realistic noise. It does not include information on false positive rates. 



\section{Conclusions and Outlook}
In order to help GP300 achieve the goal of fully autonomous detection of UHE particles with radio detection, efficient methods of signal identification are necessary. In this work, we investigated a template matching method for such signal identification. We found that a template based on realistic simulated CR signals proves to be an effective signal-identification method. It was tested using both simulated and measured background from a recently deployed GP13 detection unit.
A signal detection efficiency as high as 80\% is found for CR signals inserted in simulated and cleaned background-noise traces. This opens up the prospect of implementation on the hardware level to enable online triggering. However, this method is not very robust against strong transients that often mimic CR signals at the antenna level. Combining coincidences from more than one detector units can help reduce this issue. Furthermore, other methods of signal identification such as the combination polarization signatures of CR pulse, will be probed and validated against forthcoming measurements from various GRAND prototypes.

\section*{Acknowledgement}

This work is supported by the Polish National Agency for Academic Exchange within Polish Returns Program no. PPN/PPO/2020/1/00024/U/00001.


\bibliographystyle{unsrtnat}
\bibliography{refs}

\end{document}